\documentclass[preprint,endfloats*]{revtex4}
\nonstopmode

\usepackage{amsmath}
\usepackage{amsbsy}

\newcommand{\ket}[1]{\ensuremath{|#1\rangle}}
\newcommand{\bracket}[2]{\ensuremath{\langle #1|#2\rangle}} 
\newcommand{\dirint}[3]{\ensuremath{\langle #1|#2|#3\rangle}}

\newcommand{\bs}{\boldsymbol}

\begin{document}

\title{Reply to ``Comment on 'Modifying the variational principle in the action-integral-functional
derivation of time-dependent density functional theory' ''}
\author{J. Schirmer}
\affiliation{Theoretische Chemie, Physikalisch-Chemisches Institut,
Universit\"at Heidelberg, \\
D-69120 Heidelberg, Germany}
\date{\today}

\begin{abstract}
In a recent paper [Phys. Rev. A \textbf{77}, 062511 (2008)], it was advocated to modify
the variational principle for the action-integral functional  
in the Runge-Gross foundation of time-dependent density-functional theory.
This was criticised in a subsequent paper [Phys. Rev. A \textbf{62}, 052510 (2010)] 
by the present author.
In a Comment [Phys. Rev. A \textbf{83}, 046501 (2011)], it is argued
that the criticism is unfounded. This is a response as given 
to the four specific points raised in the Comment, clarifying and confirming       
the essence of the original criticism. 

\end{abstract}
\maketitle

In a recent paper~\cite{vig08:062511}, in the following referred to as paper I,
G. Vignale (GV) has advocated to modify
the variational principle (VP) for the action-integral functional (AIF)
in the Runge-Gross (RG) foundation of time-dependent density-functional theory (TDDFT).
This was criticised in a subsequent paper~\cite{sch10:052510} (paper II).
In a Comment~\cite{vig11:0}, GV argues that the criticism in paper II is unfounded. 
He tries to substantiate his case in four specific points, which will be
addressed below, following a differing order here.\\
\\
\noindent
\textit{Time-dependent variational principle (2):}\\
In principle, there is no disagreement here. Obviously, 
$\dirint{\delta \Psi}{i \frac{\partial}{\partial t}-\hat{H}}{\Psi} = 0$  
implies
$(i \frac{\partial}{\partial t}-\hat{H}) \ket{\Psi} =0$ if the variations
$\ket{\delta \Psi}$ are not restricted (exhaust the entire Hilbert space).
However, as has been pointed out in paper II, this usually does not apply 
to a situation where one needs to resort a time-dependent VP,
e.g., in establishing equations-of-motion (EOM) for approximate wave-functions.
In this respect, certain formulations in paper I are misleading.\\
\\
\noindent
\textit{Nature of functionals (3):}\\
In paper II (last paragraph of Sec. 4),
it is not stated ``that the wave function should be a functional not only 
of the time-dependent density itself,
but also of its derivatives with respect to time.'' Of course, the notion of
a wave-function functional in the form $\Psi(t) = \Psi[n](t)$ is ``completely
general and sufficient'', since, for a given density 
trajectory $n(t)$ (in a time interval $[t_1,t_2]$), the first and higher time derivatives are
completely determined. Nevertheless, one may ask which specific pieces of information
of the density trajectory are required in order to construct 
the wave function \textit{at a specific time} $t$ according to a given functional.

However, the remark concerning the wave-function functional is only a side note. 
The basic issue in Sec. 4 of paper II is the representation of the AIF variation $\delta A[n]$
in terms of functional derivatives.
Disregarding the phase problem for the moment,
$A[n]$ is determined completely by the density trajectory $n(\bs{r},t)$ in the considered time interval. However, this does not imply that the variation 
$\delta A[n]$ induced by a time-dependent density variation $\delta n(t)$ can be written entirely 
in terms of a functional derivative with respect to $\delta n(t)$, that is,
\begin{equation}
\delta A[n] = \int d\bs{r} \int_{t_1}^{t_2} dt \frac{\delta A[n]}{\delta n(\bs{r},t)} \delta n(\bs{r},t) 
\end{equation} 
For a given density, the functional derivative is a 
spatially and temporaly local function,
$\frac{\delta A[n]}{\delta n(\bs{r},t)} = f[n](\bs{r},t)$.
According to Eq.~(1), $\delta A[n]$ would be the sum (integral) of local contributions 
$f[n](\bs{r},t) \times \delta n(\bs{r},t)$ (times $d\bs{r} dt$)
associated with a specific point $(\bs{r},t)$ in space and time. However, it should not be taken 
for granted that 
the actual contribution to $\delta A[n]$ depends only on the static value of the
density variation at time $t$. In view of the time-derivative in the definition of $A[n]$,
one has to expect that the rate of the change of the density, that is, $d/dt \delta n(\bs{r},t)$,
will play a role as well.

As a pertinent example, one may consider the classical action $S$, briefly discussed
in Sec. 2 of paper II. Obviously, the functional $S$ is completely determined 
by the trajectory $q(t)$ in the considered time interval, that is, $S = S[q]$.
Nevertheless, the variation $\delta S$ must be written in the form
\begin{equation}
\nonumber
\delta S = \int_{t_1}^{t_2} dt \left (f(q,\dot q,t) \delta q(t) + g(q,\dot q,t) 
\delta \dot q (t) \right )
\end{equation} 
where $f = \frac{\delta S}{\delta q(t)}$ and $g = \frac{\delta S}{\delta \dot{q}(t)}$ are the (temporal) functional derivatives of $S$ with respect 
to $q(t)$ and $\dot q(t)$, respectively. (At a given \textit{point in time}, $t$, the velocity
$\delta \dot q(t)$ cannot be inferred from $\delta q(t)$, but must be supplied independently.)

It should be noted that the representation of $\delta A[n]$ according to Eq.~(1)
was introduced in the famous 1984 paper by Runge and Gross~\cite{run84:997} 
 without discussion, and has
been seen as quasi self-evident ever since. Actually, it must be seen as
the second fatal error in
the original TDDFT foundation (the first one being the 
indefiniteness of the action integral functional).\\
\\
\noindent
\textit{The ``loophole'' (4):}\\
The ``loophole'' in the argument for modifying the 
VP in paper I is not ``constructed,'' but
rather obvious: for the exact density (not to be dismissed as 
``certain particular exact densities'')
the boundary term $\bracket{\Psi[n](t_2)}{\delta \Psi[n, \delta n](t_2)}$ could vanish
due to orthogonality of $\ket{\Psi[n](t_2)}$ and 
\ket{\delta \Psi[n, \delta n](t_2)}. As long as this possibility cannot be excluded,
the case for a modified VP is not stringent.\\
\\
\noindent
\textit{Definition of the action functional (1):}\\
This is the crucial issue. Here GV writes:
``This [the finding that the RG action-integral functional is ill-defined] is false,
because multiplying the wave function by a time-dependent phase factor $e^{-i\alpha[n](t)}$,
where $\alpha[n](t)$ is an arbitrary functional of the density, $n$, and a function of time, $t$, 
amounts to adding to the Lagrangian the total time derivative
$\frac{d \alpha [n](t)}{dt}$. It is generally the case ... that the Lagrangian
is defined up to an arbitrary time derivative of a function of the coordinates and time:
it is well known that this 'gauge freedom' does not affect the \textit{variation} of the
action and therefore leaves the equations of motion unchanged.''
There are two points to be made here:\\
- Firstly, the problem with the phase function is that
$\alpha (t)$ is completely arbitrary (apart from
the possibility to fix it at the initial time $t_1$) and, in fact, cannot 
be presented as being a \textit{functional}, $\alpha[n](t)$, of the density.
The tacit assumption that the phase factor is of the form $e^{-i\alpha[n](t)}$ with
an arbitrary but otherwise well-defined functional $\alpha[n]$ amounts, logically speaking,
to a \textit{petitio principii}. I shall come back to this point below.\\
\noindent
- Secondly, an indefinitness of the Lagrangian with respect to a total time derivative
$\frac{d \alpha (t)}{dt}$ is permissible only if the variation of the boundary term 
$\alpha (t_2) - \alpha (t_1)$ in the action integral vanishes. However, this 
does not apply to the RG AIF and the original VP, since 
the variation of the phase, $\delta \alpha (t_2)$, cannot be assumed to vanish at the boundary $t_2$
of the time interval. 
(In fact, the (density induced) variation of the wave function itself does not vanish here, i.e. 
$\delta \Psi (t_2) \neq 0$, as GV rightly observes.) 
So there should be agreement that, at least in conjunction with the original VP, the RG AIF is ill-defined. 

Now let us consider the modified VP. GV claims that the 
new VP (Eq.~1 in the Comment, referred to in the following as Eq.~C-1) 
is ``completely unaffected by the arbitrary phase''.
This is demonstrated by a small derivation, of which only the resulting
Eq.~(C-2) is given in C. (For a better understanding of this argument a detailed derivation
of Eq.~(C-2) is presented in the appendix below.) The problem with this demonstration
is the unwarranted use of the arbitrary phase function as a functional of the density, $\alpha = 
\alpha[n](t)$. The derivation of Eq.~(C-2) depends manifestly on the possibility
to expand the phase function for the varied density, $n(t) + \delta[n](t)$, according
to $\alpha[n + \delta n](t) = \alpha[n](t) + \delta \alpha[n] (t)$. 
However, the actual situation is beyond remedy. Even if one assumes
that the wave function
$\Psi[n](t)$ for a specific $n(t)$ comes with a defined time-dependent phase function $\alpha(t)$, 
any phase function $\tilde{\alpha}(t)$
(with $\tilde{\alpha}(t_1) = 0$) would be permissible for the varied wave function
$\Psi[n +\delta n](t)$, so that something to the effect of a variation of the phase function, $\delta \alpha (t)$, cannot even be properly defined. This means that the indefinitness of the AIF
is not at all cured by the new VP.

A final remark concerning gauge transformations: Of course, the \textit{physics} must be invariant
with respect to gauge transformations, but what we are dealing with here is primarly
a \textit{mathematical} issue, namely an attempt to replace the time-dependent Schr\"{o}dinger equation
by an equivalent density-based EOM. Mathematically, it is manifest that a given density
determines the corresponding wave function only up to an arbitrary time-dependent phase function,
and this has mathematical implications, which should not be dismissed by referring to
a physical principle.\\
\\
\noindent
In view of the response given above to the four objections raised in the Comment,
there is no justification for the statement
that ``Schirmer's critique of my paper is invalid, and my reformulation of the variational
principle and the resolution of the causalty paradox ... stand in their pristine form.''
On the contrary, the essence of our criticism has not been rebutted:\\  
(\textit{i}) The original RG AIF is ill-defined due to the arbitrary purely time-dependent phase,
 and this problem is not eliminated in the original variational procedure.\\
(\textit{ii}) The case for a modification of the VP is not rigorous, nor is the phase problem in the definition of the AIF overcome by the modified VP.\\
(\textit{iii}) Irrespective of the phase problem, the variation of the AIF 
cannot be expressed entirely in terms of a functional derivative with respect to the density. 
It must be expected that functional derivatives with respect to the first (and possibly second)
time derivatives of the density come into play.\\

\noindent
\textbf{Appendix: Derivation of Eq.~(C-2)}\\
The idea is to compare the VP for a phase-augmented wave function functional (WFF),
$\tilde{\Psi}[n](t) = e^{-i \alpha [n] (t)} \Psi[n](t)$ 
with that for the original one, $\Psi[n](t)$.
The left-hand side of Eq.~(C-2) can readily be established, $\delta A[n]$
denoting the variation of the WFF without the additional phase. 
For the evaluation of the right-hand side of Eq.~(C-1), the starting point
is
\begin{equation}
\delta \tilde{\Psi}[n](t) = e^{-i \alpha [n + \delta n] (t)} \Psi[n + \delta n](t)
- e^{-i \alpha [n] (t)} \Psi [n](t)
\end{equation}
Using the expansions
\begin{eqnarray}
\Psi[n + \delta n](t) & = & \Psi[n](t) + \delta \Psi[n](t)\\
\alpha [n+ \delta n](t) & = & \alpha [n] (t) + \delta \alpha [n] (t)
\end{eqnarray}
gives
\begin{equation}
\delta \tilde{\Psi}[n](t) = e^{-i \alpha [n] (t)} [ e^{-i \delta \alpha [n] (t)}
(\Psi[n](t) + \delta \Psi[n](t)) - \Psi[n](t) ]
\end{equation}
Now, the scalar product on the right-hand side of Eq.~(C-1) can be evaluated
(for $t_2$) according to 
\begin{equation}
\bracket{\tilde{\Psi}[n](t_2)}{\delta \tilde{\Psi}[n](t_2)} =
e^{-i \delta \alpha [n] (t_2)} + e^{-i \delta \alpha [n] (t_2)} 
\bracket{\Psi[n](t_2)}{\delta \Psi[n](t_2)} - 1
\end{equation}
Finally, using the expansion of the exponential function,
\begin{equation}
e^{-i \delta \alpha [n] (t)} = 1 - i \delta \alpha [n] (t)
\end{equation}
the right-hand side of Eq.~(C-1) becomes (through first order in the variations)
\begin{equation}
i \bracket{\tilde{\Psi}[n](t_2)}{\delta \tilde{\Psi}[n](t_2)} = \delta \alpha[n](t_2) + i \bracket{\Psi[n](t_2)}{\delta \Psi[n](t_2)}
\end{equation}
thus completing the proof of Eq.~(C-2). The variation of the phase
drops out of Eq.~(C-2), implying that the modified VP is not affected
by adding a purely time-dependent phase factor to the wave function.
While there is nothing wrong with the calculation itself, the problem resides
in the tacitly assumed premise that the arbitrary TD phase can be treated as a functional
$\alpha[n](t)$ of the density.


\end{document}